\documentclass{article}
\usepackage{arxiv}

\usepackage{booktabs}
\usepackage{microtype}
\usepackage{cite}
\usepackage{amsmath,amssymb,amsfonts}
\usepackage{graphicx}
\usepackage{textcomp}
\usepackage{xcolor}

\def\BibTeX{{\rm B\kern-.05em{\sc i\kern-.025em b}\kern-.08em
	T\kern-.1667em\lower.7ex\hbox{E}\kern-.125emX}}

\begin{document}

\title{
	Immediate Proximity Detection Using Wi-Fi–Enabled Smartphones \\
}

\setlength{\headheight}{23pt}

\author{
	Zach Van Hyfte and Avideh Zakhor \\
	Department of Electrical Engineering and Computer Sciences \\
	University of California, Berkeley \\
	\texttt{
		\{zachvanhyfte, avz\} @berkeley.edu
	} \\
}

\maketitle

\begin{abstract}
	Smartphone apps for exposure notification and contact tracing have been shown to be effective in controlling the COVID-19 pandemic. However, Bluetooth Low Energy tokens similar to those broadcast by existing apps can still be picked up far away from the transmitting device. In this paper, we present a new class of methods for detecting whether or not two Wi-Fi–enabled devices are in immediate physical proximity, i.e. 2 or fewer meters apart, as established by the U.S. Centers for Disease Control and Prevention (CDC). Our goal is to enhance the accuracy of smartphone-based exposure notification and contact tracing systems. We present a set of binary machine learning classifiers that take as input pairs of Wi-Fi RSSI fingerprints. We empirically verify that a single classifier cannot generalize well to a range of different environments with vastly different numbers of detectable Wi-Fi Access Points (APs). However, specialized classifiers, tailored to situations where the number of detectable APs falls within a certain range, are able to detect immediate physical proximity significantly more accurately. As such, we design three classifiers for situations with low, medium, and high numbers of detectable APs. These classifiers distinguish between pairs of RSSI fingerprints recorded 2 or fewer meters apart and pairs recorded further apart but still in Bluetooth range. We characterize their balanced accuracy for this task to be between 66.8\% and 77.8\%.
\end{abstract}

\section{Introduction}

The COVID-19 pandemic sparked a proliferation of “exposure notification” or “contact tracing” smartphone apps designed to alert users if they came within close proximity of an individual infected with COVID-19 \cite{NHS-App-Impact-Paper, TraceTogether-App-Site, CA-Notify-App-Site, COVID-Watch-Arizona-App-Site}. Many were backed by public health authorities, and were used to help automate the traditional manual contact tracing process. Most of these apps continuously track, in an oblique and privacy-preserving manner, which other devices a specific smartphone has been near. When an individual tests positive for COVID-19, these apps can send a notification to every device to which that individual’s smartphone had recently been in close proximity. Those contacts can self-quarantine until they confirm that they are not carrying the virus, thus limiting the disease’s spread.

Authors in \cite{NHS-App-Impact-Paper}, studying the effects of the NHS COVID-19 app for England and Wales, estimated that “for every percentage point increase in app users, the number of cases can be reduced by [between] 0.8 … [and] 2.3 per cent … These findings provide evidence for continued development and deployment of such apps in populations that are awaiting full protection from vaccines.” COVID-19 has been shown to be much more transmissible indoors than outdoors \cite{Outdoor-COVID-Transmission-Paper, Indoor-COVID-Transmission-Paper}, and most people spend roughly 80\% of their waking hours in indoor environments. Since many public indoor environments contain a substantial number of Wi-Fi Access Points, Wi-Fi data has the potential to become a key method for exposure notification and contact tracing apps to accurately measure proximity indoors.

Currently, the majority of existing exposure notification and contact tracing apps use Bluetooth Low Energy (\textit{BLE}) to determine whether two users are in close proximity. Specifically, a device running an exposure notification app periodically broadcasts, via BLE, a token that can be picked up by other nearby devices running the same app. The distance from which a transmitted BLE token similar to those broadcast by the apps can be received varies based on the radio configuration and the characteristics of the surrounding environment, but can be over 20 meters.

An active research area over the past 20 years has been the development of systems for localizing Wi-Fi–enabled devices in indoor environments using a Wi-Fi RSSI fingerprint, which is a list of the Wi-Fi Access Points (\textit{APs}) detected by a particular device, and the received signal strength indicator (\textit{RSSI}) value for each of those APs \cite{Indoor-Positioning-Survey-Paper}. Existing Wi-Fi localization systems can determine a device’s position within a building with an average error of several meters.

In this paper, we present a new class of methods for detecting whether or not two Wi-Fi–enabled devices are in immediate physical proximity, i.e. 2 or fewer meters apart, as established by the U.S. Centers for Disease Control and Prevention (CDC). Each classifier takes as input two Wi-Fi RSSI fingerprints — two sets of MAC addresses of \textit{APs} and their respective received signal strengths — and predicts whether or not the two fingerprints were captured within roughly two meters of each other. These methods can work across a wide range of Wi-Fi–enabled device types, and are designed to be more robust than earlier Wi-Fi–based methods in dealing with a wide range of physical environments as well as heterogeneous devices. Methods such as the ones we describe could supplant or augment existing Bluetooth-based proximity detection methods employed by exposure notification apps. They could improve the accuracy of smartphone-based contact tracing by determining which individuals were actually in extremely close physical proximity to an infected individual.

The outline of this paper is as follows: In Section \ref{related-work-section}, we discuss how our methods differ from previous work. In Section \ref{data-sources-section}, we outline where the RSSI fingerprint data used in our experiments was sourced. In Section \ref{proximity-classes-section}, we describe how our training and evaluation sets were created from the RSSI fingerprint data. In Section \ref{feature-extraction-section}, we discuss each of the fingerprint similarity measurements that are passed to our classifiers as input features. In Section \ref{mitigating-device-heterogeneity-section}, we detail the additions and adjustments we made to the input features to ensure that our classifiers work across a heterogeneous array of devices. In Section \ref{experiment-setup-section}, we describe the algorithms, software, and settings used for training the classifiers. In Section \ref{generic-classifier-section}, we present empirical evidence indicating that a single machine learning classifier cannot generalize well to a range of different environments with vastly different numbers of detectable APs. In \ref{specialized-classifiers-section}, we present a set of three binary machine learning classifiers, each tailored for a situation where the number of APs detected by the user falls into a given range, and quantify their performance. Finally, in Section \ref{conclusions-future-work-section}, we conclude the paper.

\section{Related Work}
\label{related-work-section}

The construction of indoor localization systems based on RSSI fingerprints for use within a specific building has been an active research area for over a decade \cite{Indoor-Positioning-Survey-Paper}. These systems typically require the creation of a database containing at least one RSSI fingerprint from every single location or position within the building. Wi-Fi–enabled devices within the building can submit a fingerprint to a central server, which employs an algorithm such as $k$-nearest neighbors to find the location in the database whose fingerprint is most similar to the one submitted by the device. The classifiers we present are concerned with determining whether or not two users are in close physical proximity, rather than estimating a user’s location on a map, but they make use of established techniques designed to increase the accuracy of RSSI fingerprint–based localization systems. Crucially, too, the proximity detection methods explored in this paper do not require the \textit{a priori} collection of a database of fingerprints from every possible reference point.

Similar to this paper, \cite{Inferring-WiFi-Paper} presents a number of binary machine learning classifiers that use various features extracted from a pair of Wi-Fi fingerprints to determine whether or not they were recorded in “close physical proximity” to each other. However, the sole criterion used in \cite{Inferring-WiFi-Paper} to determine whether two fingerprints were classified as within close physical proximity was whether the two smartphones recording the fingerprints were both able to detect each other via Bluetooth around the time the two fingerprints were recorded; this, it was noted, generally happens when the two phones are 10 or fewer meters away from each other. In contrast, because the data used in our training and evaluation sets includes precise, meter-level ground truth position information, we assign a label of “Close” only to samples recorded within 2.25 meters of each other — the goal of our classifiers is to determine whether two devices are not just nearby but in \textit{immediate} physical proximity to each other, i.e. less than the safe social distance defined by the U.S. Centers for Disease Control and Prevention (CDC). In addition, our systems expand upon the set of similarity measurements evaluated in \cite{Inferring-WiFi-Paper}, making use of additional features and pre-processing steps to further enhance accuracy.

\section{Data Sources}
\label{data-sources-section}

A variety of Wi-Fi fingerprint datasets used to develop and evaluate Wi-Fi localization systems are publicly available online, as shown in Table \ref{wifi-fingerprint-datasets}; we used Wi-Fi fingerprints from several of these datasets to create training and evaluation data for our classifiers.

\begin{table}[bt]

	\caption{Wi-Fi Fingerprint Datasets}

	\begin{center}
	\begin{tabular}{ccc}
		\toprule
			\textbf{Name} & $\textbf{Median APs}^\mathrm{a}$ & \textbf{Source} \\
		\midrule
			Miskolc & $10$ & UCI ML Repository \cite{Miskolc-Dataset-Paper} \\
			$\text{JUIndoorLoc}^\mathrm{b}$ & $15$ & JUIndoorLoc Paper \cite{JUIndoorLoc-Dataset-Paper} \\
			UJIndoorLoc & $17$ & UJI IndoorLoc Platform \cite{IndoorLoc-Platform-Site} \\
			IPIN 2016 Tutorial & $32$ & UJI IndoorLoc Platform \cite{IndoorLoc-Platform-Site} \\
			TampereU & $38$ & UJI IndoorLoc Platform \cite{IndoorLoc-Platform-Site} \\
			Alcalá Tutorial & $40$ & UJI IndoorLoc Platform \cite{IndoorLoc-Platform-Site} \\
			Suburban Home & $8$ & Collected by Us \\
			Cory Hall & $66$ & Collected by Us \\
		\bottomrule \\
			\multicolumn{3}{l}{$^{\mathrm{a}}$ The median number of APs detected in a given fingerprint} \\
			\multicolumn{3}{l}{from the dataset.} \\
			\multicolumn{3}{l}{$^{\mathrm{b}}$ Only the test data from this dataset was used.}
	\end{tabular}
	\label{wifi-fingerprint-datasets}
	\end{center}

\end{table}

To the extent possible, given state and local COVID-19 restrictions and shelter-in-place orders, we also collected our own training and evaluation data, optimized for training proximity detection classifiers. We developed an Android app for data collection, which uses Android’s \texttt{WifiManager} API to perform scans of nearby APs on-demand and save the fingerprints to the device for later use.

Once certain buildings on the UC Berkeley campus re-opened for limited research activity, we were able to collect our own RSSI fingerprint data in Cory Hall, a five-story academic building on the UC Berkeley campus that houses classrooms, labs, common work spaces, and offices for the EECS department. We collected a total of $738$, $1521$, and $603$ fingerprints from the first, third, and fourth floors of the building, respectively, over the course of three days in late November and early December 2020. Fingerprints were collected with the following devices: (a) a Google Pixel 3 running Android 10: \textit{$225$, $576$, and $198$ fingerprints}; (b) a Nokia 2.2 \footnote[1]{Model TA-1179, in the 3 GB RAM configuration}, running Android 10: \textit{$225$, $576$, and $198$ fingerprints}; (c) a first-generation Google Pixel XL running Android 10: \textit{$225$, $342$, and $198$ fingerprints}; (d) an Oppo RX17 Pro running Android 8.1: \textit{$63$, $27$, and $9$ fingerprints}.

For each of the devices listed above, the three numbers reflect the number of fingerprints recorded with the device on the first, third and fourth floors of Cory Hall, respectively. At each position where a set of fingerprints were recorded, a Leica DISTO E7100i laser distance measurement device was used to measure the distance from each point to two reference walls on the current floor, which were later used to calculate the distance between samples from the same floor. Each device captured data in “bursts,” recording $9$ fingerprints at a time in rapid succession. The median number of APs observed in a given fingerprint from the first, third, and fourth floors are $76$, $69$, and $52$, respectively. The median number of APs observed in a given fingerprint from the entire set of collected data is $66$.

Data was also collected from a single-story suburban home in San Diego, California, over an area spanning roughly 1,800 square feet, also using our Android data collection app. Fingerprints were recorded with a Google Pixel 3 and a Nokia 2.2 \footnotemark[1], both running Android 10. At $63$ different locations within the home, each device recorded a “burst” of $9$ fingerprints, for a total of $1,134$ unique fingerprints. Within the home itself are two APs, one located roughly in the northwest corner of the floorplan, and the other roughly in the southeast corner of the floorplan. However, at any given location within the home, APs from several nearby homes are also detectable; across all of the samples collected, the median number of APs observed in a given fingerprint is $8$.

\section{Proximity Classes}
\label{proximity-classes-section}

Our classifiers evaluate two RSSI fingerprints at a time, determining whether or not they were recorded in immediate physical proximity to each other. To prepare training and evaluation sets from a given pool of fingerprints, we isolate fingerprints into different subsets, one for each floor of each building of each dataset. Within each subset, we enumerate every possible pairing of two distinct fingerprints from the subset, and calculate the two-dimensional distance $d$ between the locations where the two fingerprints were recorded, using the meter-level coordinates that are either provided in the dataset or recorded by our team. We assigned each pairing of fingerprints a “proximity class” according to the value of $d$ for that pairing. If $0 \hspace{0.1cm}\text{meters} \le d \le 2.25 \hspace{0.1cm}\text{meters}$, the fingerprint pairing’s proximity class is set to “Close.” If $3.25 \hspace{0.1cm}\text{meters} \le d \le 20 \hspace{0.1cm}\text{meters}$, it is set to “Far.”

Fingerprint pairings are dropped from the training or evaluation sets if the two fingerprints were recorded more than 20 meters apart, in an effort to focus the training process on the fine-grained differentiation between pairings recorded in immediate physical proximity and those recorded in somewhat close physical proximity, i.e. Bluetooth range. This is in contrast to an approach which would optimize our classifiers for the coarse-grained differentiation of pairings recorded in immediate physical proximity and those recorded very far away from each other. Fingerprint pairings are also dropped from the training or evaluation data if the two fingerprints were recorded between $2.25$ and $3.25$ meters apart, in an effort to focus the training process on differentiating between pairings recorded in immediate physical proximity and those recorded in somewhat close physical proximity, rather than on border cases that are nearly at the distance cutoff for the “Close” label.

\section{Classifier Input Features}
\label{feature-extraction-section}

For each fingerprint pairing, we calculate an expansive set of features, which are the inputs passed directly to the classifiers described in Sections \ref{generic-classifier-section} and \ref{specialized-classifiers-section}. The majority of these input features were selected because we hypothesized that they could potentially serve as similarity measurements — quantities that numerically express the level of quantitative similarity between the two fingerprints in a pairing, which can be a rough proxy for how close together the fingerprints were recorded. In this section, we describe our chosen features in detail. Consider a pair of RSSI fingerprints, $F_X$ and $F_Y$. Define the set of \textit{shared APs} $\{S_1$, $S_2$, …, $S_N\}$ as the APs that are detected in both $F_X$ and $F_Y$. Furthermore, define $\text{RSSI}(a, F_X)$ and $\text{RSSI}(a, F_Y)$ as the RSSI of the AP $a$ in fingerprints $F_X$ and $F_Y$, respectively.

\subsection{AP Detection–Based Features}

The classifiers are provided with the following input features derived from the number of APs detected in $F_X$, the number detected in $F_Y$, and the number of shared APs:

\begin{enumerate}
    \item The \textit{shared AP count}: $N$, the number of shared APs from above.
    \item The \textit{union AP count}: the total number of APs detected in at least one of the fingerprints.
    \item The \textit{non-shared AP count}: the total number of APs detected in exactly one of the two fingerprints, but not both.
    \item The \textit{detected AP count difference}: the absolute value of the difference between the number of the APs detected in $F_X$ and the number of APs detected in $F_Y$.
    \item The \textit{Jaccard similarity} of the sets of APs detected in the two fingerprints: the shared AP count divided by the union AP count.
\end{enumerate}

\subsection{Basic RSSI Value–Based Features}

The \textit{Manhattan distance} and the \textit{Euclidean distance} between two fingerprints $F_X$ and $F_Y$, calcualted exactly as in \cite{Inferring-WiFi-Paper}, are both input to the classifiers.

Define the \textit{top AP(s)} of a fingerprint as the AP(s) whose measured RSSI value(s) in that fingerprint are the highest among all APs detected in that fingerprint. We define a number of feature types related to the difference in dBm between the shared APs’ measured RSSI values in $F_X$ and in $F_Y$:

\begin{enumerate}

	\item The feature \textit{Has shared top AP within $Z$ dBm} for fingerprints $F_X$ and $F_Y$ is equal to $1$ if there exists at least one shared AP $S_i$ such that (a) the measured RSSI value of $S_i$ in $F_X$ is at most $Z$ dBm below the maximum RSSI of any AP in $F_X$, and (b) the measured RSSI value of $S_i$ in $F_Y$ is at most $Z$ dBm below the maximum RSSI of any AP in $F_Y$. Otherwise, the feature is equal to $0$. Features of the type \textit{Has shared top AP within Z dBm} for $Z = 1$, $2$, …, $15$ were provided to the classifiers. This feature type is similar to the “top AP $\pm$ 6 dB” feature described in \cite{Inferring-WiFi-Paper}. It is designed to allow a classifier to determine whether there is an AP that both fingerprints were likely closer to than most other APs, while accounting for differences in RSSI measurement scales or minor RSSI fluctuations that could change which specific AP has the highest RSSI.

	\item The feature \textit{RSSIs within $Z$ dBm percentage} for fingerprints $F_X$ and $F_Y$ is equal to the percentage of shared APs whose RSSI values in $F_X$ are within $Z$ dBm of their RSSI values in $F_Y$. Features of the type \textit{RSSIs within $Z$ dBm percentage} were provided to the classifiers for $Z = 1$, $2$, …, $15$.

	\item The feature \textit{Has shared top $K$ APs} for fingerprints $F_X$ and $F_Y$ is equal to $1$ if and only if the $K$ highest-RSSI APs in $F_X$ are the same as the $K$ highest-RSSI APs in $F_Y$, regardless of ordering differences between the two fingerprints. Features of the type \textit{Has shared top $K$ APs} were provided to the classifiers for $K = 1$, $2$, …, $8$.

\end{enumerate}

\subsection{Redpin Score–Based Features}

Two input features based on the Redpin score are also provided to the classifiers. The Redpin score is a measurement of fingerprint similarity that was developed for and used in the Redpin crowdsourced Wi-Fi localization system \cite{Redpin-Paper}, with a reference implementation available at \cite{Redpin-Code-Repository}. The Redpin score calculation is not a commutative operation, so we provide two Redpin scores to the classifiers: $\texttt{RedpinScore}(\max(F_X, F_Y), \min(F_X, F_Y))$ and $\texttt{RedpinScore}(\min(F_X, F_Y), \max(F_X, F_Y))$, where $\min$ and $\max$ select the fingerprint with the lower and higher number of detected APs, respectively.

\subsection{Correlation–Based Features}
\label{correlation-features-subsection}

In addition, we define the following pairs of vectors:

\begin{enumerate}

	\item The \textit{shared AP RSSI value vectors} are a pair of vectors containing the measured RSSI of each shared AP in $F_X$ and in $F_Y$:
		$$ \Big[ \hspace{0.1cm} \text{RSSI}(S_1, F_X), \hspace{0.1cm} \text{RSSI}(S_2, F_X), \hspace{0.1cm} \ldots \hspace{0.1cm} \text{RSSI}(S_N, F_X) \hspace{0.1cm} \Big] $$
		$$ \Big[ \hspace{0.1cm} \text{RSSI}(S_1, F_X), \hspace{0.1cm} \text{RSSI}(S_2, F_X), \hspace{0.1cm} \ldots \hspace{0.1cm} \text{RSSI}(S_N, F_X) \hspace{0.1cm} \Big] $$

	\item The \textit{shared AP pair difference vectors} are a pair of vectors containing the absolute value of the difference between the RSSI values within each fingerprint for every possible pairing of two distinct shared APs:
		$$ \Big[ \hspace{0.1cm} \lvert \hspace{0.1cm} \text{RSSI}(S_i, F_X) \hspace{0.1cm} - \hspace{0.1cm} \text{RSSI}(S_j, F_X) \hspace{0.1cm} \rvert \hspace{0.2cm} \ldots \hspace{0.2cm} \forall (i \neq j) \le N \hspace{0.1cm} \Big] $$
		$$ \Big[ \hspace{0.1cm} \lvert \hspace{0.1cm} \text{RSSI}(S_i, F_Y) \hspace{0.1cm} - \hspace{0.1cm} \text{RSSI}(S_j, F_Y) \hspace{0.1cm} \rvert \hspace{0.2cm} \ldots \hspace{0.2cm} \forall (i \neq j) \le N \hspace{0.1cm} \Big] $$

	\item The \textit{shared AP pair ratio vectors} are a pair of vectors containing the ratios of the RSSI values within each fingerprint of every possible pairing of two distinct shared APs. They are similar to the modified type of RSSI fingerprints used in Hyperbolic Location Fingerprinting \cite{HLF-Pair-Ratio-Paper}:
		$$ \Big[ \hspace{0.1cm} \frac{\text{RSSI}(S_i, F_X)}{ \text{RSSI}(S_j, F_X)} \hspace{0.2cm} \ldots \hspace{0.2cm} \forall (i, j) \le N, \hspace{0.1cm} i \neq j \hspace{0.1cm} \Big] $$
		$$ \Big[ \hspace{0.1cm} \frac{\text{RSSI}(S_i, F_Y)}{ \text{RSSI}(S_j, F_Y)} \hspace{0.2cm} \ldots \hspace{0.2cm} \forall (i, j) \le N, \hspace{0.1cm} i \neq j \hspace{0.1cm} \Big] $$

	\item Define the \textit{rank} of a shared AP $S_i$ within a particular fingerprint as the number of shared APs, including $S_i$ itself, in the fingerprint whose measured RSSI values are at least as weak as $S_i$’s. The \textit{normalized ordered shared AP rank vectors} are a pair of vectors that are both normalized to unit vectors. One is always the normalized form of:
		$$ R_X = \big[ \hspace{0.1cm} N, \hspace{0.1cm} N - 1, \hspace{0.1cm} N - 2, \hspace{0.2cm} \ldots \hspace{0.2cm} \hspace{0.1cm} 2, \hspace{0.1cm} 1 \hspace{0.1cm} \big] $$.
	The other is the normalized form of:
		$$ R_Y = \big[ \hspace{0.1cm} \text{Rank}_Y(N), \hspace{0.1cm} \text{Rank}_Y(N - 1), \hspace{0.2cm} \ldots \hspace{0.2cm} \hspace{0.1cm} \text{Rank}_Y(1) \hspace{0.1cm} \big] $$
	$\text{Rank}_Y(i)$, as used above, is the rank within fingerprint $F_Y$ of the shared AP with rank $i$ in fingerprint $F_X$. Thus, any index $i$ represents some shared AP, and $R_X[i]$ and $R_Y[i]$ are the ranks of that shared AP in the two different fingerprints. The correlation coefficient of $R_X$ and $R_Y$ therefore measures how similar the signal strength rankings of the shared APs are across the two fingerprints.

\end{enumerate}

For each of the pairs of vectors above, namely, the shared AP RSSI vectors, the shared AP pair difference vectors, the shared AP pair ratio vectors, and the normalized shared AP rank vectors, the following measurements of similarity between the two vectors are passed as input features to the classifiers: (a) Cosine similarity, (b) Pearson coefficient, (c) Spearman coefficient, and (d) Kendall coefficient.

\subsection{Difference-Based Features}

Lastly, we define a second set of individual vectors as follows:

\begin{enumerate}

	\item The \textit{shared AP RSSI difference vector} is a vector containing the absolute value of the difference between the measured RSSI in $F_X$ and in $F_Y$ of each shared AP:
		$$ \Big[ \hspace{0.1cm} \lvert \hspace{0.1cm} \text{RSSI}(S_i, \hspace{0.05cm} F_X) \hspace{0.1cm} - \hspace{0.1cm} \text{RSSI}(S_i, \hspace{0.05cm} F_Y) \hspace{0.1cm} \rvert \hspace{0.2cm} \ldots \hspace{0.2cm} \forall \hspace{0.1cm} i \le N \hspace{0.1cm} \Big] $$

	\item Define $\text{PD}(S_i, \hspace{0.1cm} S_j, \hspace{0.1cm} F_X)$ and $\text{PD}(S_i, \hspace{0.1cm} S_j, \hspace{0.1cm} F_Y)$ as the shared AP pair difference, as defined in Section \ref{correlation-features-subsection}, of shared APs $S_i$ and $S_j$ in fingerprints $F_X$ and $F_Y$ respectively. The \textit{shared AP pair difference comparison vector} contains the absolute value of the difference between the pair difference in $F_X$ and the pair difference in $F_Y$ of every possible pairing of two distinct shared APs:
		$$ \Big[ \hspace{0.1cm} \big\lvert \hspace{0.1cm} \text{PD}(S_i, \hspace{0.1cm} S_j, \hspace{0.1cm} F_X) \hspace{0.1cm} - \hspace{0.1cm} \text{PD}(S_i, \hspace{0.1cm} S_j, \hspace{0.1cm} F_Y) \hspace{0.1cm} \big\rvert \hspace{0.2cm} \ldots \hspace{0.2cm} \forall \hspace{0.1cm} i \le N \hspace{0.1cm} \Big] $$

	\item Similarly, define $\text{PR}(S_i, S_j, F_X)$ and $\text{PR}(S_i, S_j, F_Y)$ as the shared AP pair ratio, as defined above, of shared APs $S_i$ and $S_j$ in fingerprints $F_X$ and $F_Y$ respectively, and define the \textit{shared AP pair ratio comparison vector} as follows:
		$$ \Big[ \hspace{0.1cm} \big\lvert \hspace{0.1cm} \text{PR}(S_i, \hspace{0.1cm} S_j, \hspace{0.1cm} F_X) \hspace{0.1cm} - \hspace{0.1cm} \text{PR}(S_i, \hspace{0.1cm} S_j, \hspace{0.1cm} F_Y) \hspace{0.1cm} \big\rvert \hspace{0.2cm} \ldots \hspace{0.2cm} \forall \hspace{0.1cm} i \le N \hspace{0.1cm} \Big] $$

\end{enumerate}

For each of the individual vectors above, namely, the shared AP RSSI difference vector, the shared AP pair difference comparison vector, and the shared AP pair ratio comparison vector, the following similarity measurements are passed as input features to the classifiers: (a) the smallest element of the vector; (b) the largest element of the vector; (c) the mean of all of the vector’s elements; (d) the median of all of the vector’s elements; (e) the harmonic mean of all of the vector’s elements; (f) the standard deviation of all of the vector’s elements; and (g) the population standard deviation of all of the vector’s elements.

\section{Mitigating the Effects of Device Heterogeneity}
\label{mitigating-device-heterogeneity-section}

A common challenge in creating indoor positioning systems that use RSSI fingerprints is dealing with the effects of device heterogeneity. Since different devices and different Wi-Fi chips have different levels of sensitivity and different signal strength measurement scales, two RSSI fingerprints recorded in the same place with different devices can generally have (a) less of an overlap in the set of APs detected, and (b) a larger difference between the two fingerprints’ measured RSSI values for shared APs than a pair of RSSI fingerprints recorded with the same device model. We employ several measures to compensate for device heterogeneity, including several additional features input to the classifier:

\begin{enumerate}

	\item In both publicly available datasets and in our own collected data, among the metadata stored with each fingerprint is the model of the device that recorded it. Therefore, an additional feature passed into all of our classifiers is \textit{Identical recording devices}, which is $1$ if the two fingerprints were recorded by the same device model and $0$ otherwise. This additional feature allows a classifier to select different decision sequences or feature weights and cutoff values for device-heterogeneous fingerprint pairings than for device-homogeneous ones.

	\item An additional similarity measurement, designed to be more robust in the face of inputs from heterogeneous devices, is the Refined Relative RSSI Relationship (\textit{RE3}), as proposed for use in Wi-Fi localization systems in \cite{RE3-Paper}. Since it is calculated using the ranks of the APs detected in a fingerprint — namely, the AP with the highest RSSI has the top rank and the one with the lowest RSSI has the lowest — rather than their explicit RSSI values, it is more likely to remain the same across different devices that have different RSSI measurement scales. In addition to all of the classifier input features described above, the RE3 of the two fingerprints is calculated and passed as an input feature to the classifier.

	\item It is established that in many cases, given two fingerprints recorded in the same location with different devices, the measured RSSI values in both fingerprints are roughly linearly dependent. Applying a linear transformation to the RSSI values from one of the fingerprints can bring the two fingerprints in line and compensate for differences in the two devices’ sensitivities and signal strength measurement scales. So, in actuality, for each of the 80 classifier input features described above that depends on the APs’ RSSI values themselves, four separate input features are passed into the classifiers:

		\begin{enumerate}

			\item \textit{No transformation}: The feature value as calculated with the raw, original RSSI values from $F_X$ and $F_Y$.

			\item \textit{Single-fingerprint least squares}: The feature value as calculated after every RSSI value $r_X$ in the fingerprint $F_X$ has been replaced with $Ar_X + B$, with the values of the constants $A$ and $B$ determined by using the method of least squares to fit the RSSI values from $F_X$ to the RSSI values from $F_Y$.

			\item \textit{Single-fingerprint 50\% least squares}: The feature value as calculated after every RSSI value $r_X$ in the fingerprint $F_X$ has been replaced with $\frac{A}{2}r_X + \frac{B}{2}$, where $A$ and $B$ are computed as in (b) above.

			\item \textit{Double-fingerprint least squares}: The feature value as calculated after every RSSI value $r_X$ in the fingerprint $F_X$ has been replaced with $Ar_X + B$ and every RSSI value $r_Y$ in the fingerprint $F_Y$ has been replaced with $Cr_Y + D$, where $A$ and $B$ are the same as in (b) and (c) above, and $C$ and $D$ are similarly determined by using the method of least squares to fit the original, raw RSSI values from $F_Y$ to the original, raw RSSI values from $F_X$.

		\end{enumerate}

\end{enumerate}

\section{Experimental Setup}
\label{experiment-setup-section}

We now present a series of binary machine learning classifiers whose aim is to predict, based on the extensive set of input features they are provided, the proximity class — “Close” or “Far” — of input fingerprint pairings.

The high number of input features provided to the classifiers has the potential to make our data subject to overfitting and the “curse of dimensionality,” summarized in \cite{Ensemble-Survey-Paper} as the phenomenon where “increasing the number of features fed into a machine learning model usually exponentially increases the search space and hence, the probability of fitting models that cannot be generalized.” To mitigate this, while preserving the ability to take advantage of a large set of available features, we trained classifiers using the attribute bagging method, introduced in \cite{Attribute-Bagging-Paper}. Attribute bagging is “a wrapper method that can be used with any learning algorithm,” in which many different subsets of features, usually small ones, are randomly selected and used to train a set of smaller base estimators, usually decision trees. Base estimators with the highest possible performance are combined into an ensemble classifier. When a sample is input to an attribute bagging classifier for prediction, the base estimators each individually predict, or “vote,” based on their particular set of input features, which class a sample belongs to. The class predicted by the most base estimators, after optionally factoring in weights assigned to each base estimator’s vote, is chosen as the final predicted class. Scikit-learn \cite{Scikit-learn-Paper}, a popular Python machine learning library, provides a flexible \texttt{BaggingClassifier} class that can be configured to perform attribute bagging. All of the classifiers outlined below were trained with instances of the \texttt{BaggingClassifier} class from scikit-learn 0.22.1, with \texttt{DecisionTreeClassifier} instances as the base estimators. Through manual tuning across a number of experiments, we empirically found a set of hyperparameters that generally work well for this problem: at most $3$ features per base estimator, and $300$ individual voting base estimators, i.e. \texttt{max\_features = 3} and \texttt{n\_estimators = 300}.

To further reduce the dimensionality of the data, and reduce both the probability of overfitting and the time it takes to train a classifier, when necessary, we used the minimum Redundancy Maximum Relevance (\textit{mRMR}) feature selection algorithm presented in \cite{mRMR-Paper}. The mRMR algorithm analyzes a dataset and aims to identify a set of features that are both “maximally relevant,” i.e. that “have the largest mutual information…with the target class,” and “minimally redundant,” i.e. that have the smallest amount of mutual information amongst each other. We used version 0.1.11 of the pymRMR Python package provided by the authors of \cite{mRMR-Paper}, with the “mutual information difference” (\textit{MID}) feature selection method, to select the top features.

For every experiment, we began by training on a perfectly class-balanced training set, where exactly 50\% of the samples in the training set belong to the “Close” class and the remaining 50\% belong to the “Far” class. However, in some cases, training on a perfectly class-balanced training set produced a classifier that was overly biased towards one particular class. In the experiments discussed below, when applicable, we adjusted the class balance of the training set through manual iteration to compensate for any biases and produce a classifier with better overall performance.

We began by training and evaluating a single classifier on data from a wide variety of different environments. After verifying that the generic classifier approach was infeasible, we developed specialized classifiers for use with fingerprints containing different numbers of detected APs.

\section{Generic Classifier Experiments}
\label{generic-classifier-section}

Initially, we trained a single generic classifier on a combination of data from all of the publicly available Wi-Fi localization datasets listed in Table \ref{wifi-fingerprint-datasets}. Table \ref{number-of-samples-in-datasets} shows the number of usable samples yielded by the training set generation process described in Sections \ref{proximity-classes-section} and \ref{feature-extraction-section} for each dataset.

\begin{table}[bt]

	\caption{Number of Fingerprint Pairings Generated from Individual Datasets}

	\begin{center}
	\begin{tabular}{ccc}
		\toprule
			\textbf{Dataset Name} & \textbf{“Far” Samples} & \textbf{“Close” Samples} \\
		\midrule
			Miskolc & $227,716$ & $11,630$ \\
			JUIndoorLoc & $554,011$ & $335,002$ \\
			UJIndoorLoc & $2,644,089$ & $387,186$ \\
			IPIN 2016 Tutorial & $90,964$ & $30,845$ \\
			TampereU & $374,556$ & $17,193$ \\
			Alcalá Tutorial & $826,009$ & $110,442$ \\
			Suburban Home & $491,832$ & $97,443$ \\
			Cory Hall, Floor 1 & $107,892$ & $24,336$ \\
			Cory Hall, Floor 3 & $314,847$ & $34,029$ \\
			Cory Hall, Floor 4 & $65,124$ & $16,263$ \\
		\bottomrule
	\end{tabular}
	\label{number-of-samples-in-datasets}
	\end{center}

\end{table}

We created a training set with samples evenly distributed among the datasets listed in Table \ref{wifi-fingerprint-datasets} by randomly selecting $9,000$ “Close” and $8,000$ “Far” fingerprint pairings from each of the datasets. We trained a \texttt{BaggingClassifier} on this training set, and individually evaluated the classifier on the remaining samples, i.e. those not selected for use in the training set, from each dataset. The percentage of true negative samples correctly identified, the percentage of true positive samples correctly identified, and the balanced accuracy of the classifier for these “evaluation” portions of all of the datasets are shown in Table \ref{generic-classifier-experiment-results}. The generic classifier’s performance is not only low but inconsistent, varying across different environments even when data from each of those environments is present in the training set.

\begin{table}[bt]

	\caption{Experiment Results for Generic Classifier on Individual Datasets}

	\begin{center}
	\begin{tabular}{cccc}
		\toprule
			\textbf{Dataset Name} & \textbf{True Positives} & \textbf{True Negatives} & \textbf{Balanced Accuracy} \\
		\midrule
			Miskolc & $75.76\%$ & $42.47\%$ & $59.11\%$ \\
			JUIndoorLoc & $36.00\%$ & $68.69\%$ & $52.34\%$ \\
			UJIndoorLoc & $64.30\%$ & $57.82\%$ & $61.06\%$ \\
			IPIN 2016 Tutorial & $71.71\%$ & $39.21\%$ & $55.46\%$ \\
			TampereU & $75.92\%$ & $42.14\%$ & $59.03\%$ \\
			Alcalá Tutorial & $84.48\%$ & $44.59\%$ & $64.53\%$ \\
			Suburban Home & $42.75\%$ & $72.74\%$ & $57.74\%$ \\
			Cory Hall & $59.06\%$ & $70.09\%$ & $63.57\%$ \\
		\midrule
			\textcolor{darkgray}{Average} & $63.75\%$ & $54.72\%$ & $59.27\%$ \\
		\bottomrule
	\end{tabular}
	\label{generic-classifier-experiment-results}
	\end{center}

\end{table}

The results of the the generic classifier experiment indicate that a single, one-size-fits all classifier is insufficient for providing consistent performance across a wide range of environments with different average numbers of APs per fingerprint.

\section{Specialized-Classifier Experiments}
\label{specialized-classifiers-section}

To pursue solid performance across different environments, we developed an ensemble of three classifiers, each tailored specifically for environments with a particular “AP density,” or number of APs typically detected per fingerprint. We developed specialized classifiers for three different types of target environments: (a) locations with a low AP density — $5$ to $15$ APs detected per fingerprint on average; (b) locations with a moderate to high AP density — $30$ to $70$ APs detected per fingerprint on average; and (c) locations with a very high AP density — $70$ to $90$ APs detected per fingerprint on average.

\begin{figure}[bt]
	\centerline{\includegraphics[scale=0.40]{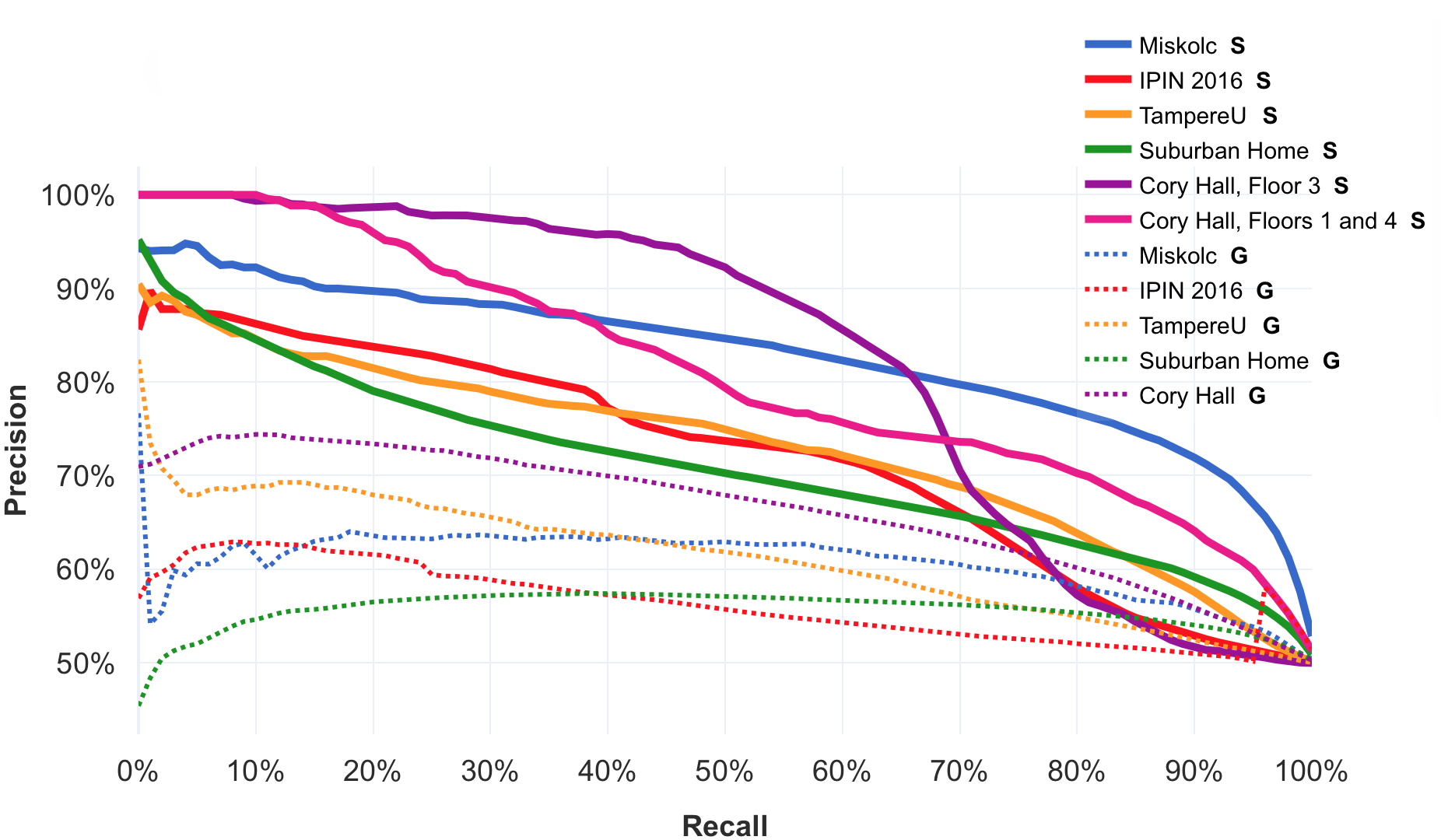}}
	\caption{
		Precision–recall curves of the specialized classifiers on perfectly class-balanced subsets of each evaluation dataset from \ref{specialized-classifiers-experiment-results}, along with the analogous precision–recall curves of the generic classifier. \\
		$S$ = specialized classifier, $G$ = generic classifier
	}
	\label{generic-vs-specialized-classifier-precision-recall-curves}
\end{figure}

Table \ref{specialized-classifiers-experiment-results} shows the percentage of true negative samples correctly identified, the percentage of true positive samples correctly identified, and the balanced accuracy for all of the experiments ran with the specialized classifiers. Figure \ref{generic-vs-specialized-classifier-precision-recall-curves} shows the precision–recall curves of the specialized classifiers on perfectly class-balanced subsets of each evaluation set, along with the analogous precision–recall curves of the generic classifier on perfectly class-balanced subsets of the evaluation sets described in Section \ref{generic-classifier-section}.

In what follows, we describe the details of the specialized classifiers developed for each of the three types of target environments above.

\begin{table}[bt]

	\caption{Experiment Results for Specialized Classifiers}

	\begin{center}
	\begin{tabular}{cccccc}
		\toprule
			\textbf{\#} & \textbf{Dataset Name (Features)} & \textbf{Classifier} & \textbf{True Negatives} & \textbf{True Positives} & \textbf{Balanced Accuracy} \\
		\midrule
			1 & Miskolc (All) & Low AP & $71.66\%$ & $84.12\%$ & $77.89\%$ \\
			2 & Suburban H. (All) & Low AP & $63.45\%$ & $70.25\%$ & $66.85\%$ \\
			3 & Miskolc (Top 7) & Low AP & $71.31\%$ & $78.53\%$ & $74.92\%$ \\
		\midrule
			4 & TampereU (All) & Medium AP & $70.77\%$ & $68.06\%$ & $69.41\%$ \\
			5 & IPIN 2016 (All) & Medium AP & $69.62\%$ & $66.02\%$ & $67.82\%$ \\
		\midrule
			6 & Cory Hall, Floor 3 (All) & High AP & $69.26\%$ & $70.43\%$ & $69.84\%$ \\
			7 & Cory Hall, Floors 1 and 4 (All) & High AP & $70.38\%$ & $73.47\%$ & $71.92\%$ \\
		\bottomrule
	\end{tabular}
	\label{specialized-classifiers-experiment-results}
	\end{center}

\end{table}

\subsection{Low AP Density Classifier}

The median number of APs detected in a fingerprint from the data collected from the suburban home and in a fingerprint from the publicly available Miskolc dataset is $8$ and $10$, respectively, making them both low AP density environments. Using the process described in Section \ref{experiment-setup-section}, we created a training set from the suburban home data containing all $97,443$ “Close” and $97,443$ randomly chosen “Far” fingerprint pairings, and used it to train a \texttt{BaggingClassifier}. We then evaluated the performance of the classifier on the entire set of fingerprint pairings extracted from the publicly available Miskolc dataset. The results of this evaluation, shown in row 1 of Table \ref{specialized-classifiers-experiment-results}, indicate that a model trained on data from a single location can achieve fairly high recall on data from a completely different location with a similar AP density.

For comparison, we swapped the training and evaluation datasets, creating a class-imbalanced training set with all $11,630$ “Close” and $8,000$ randomly selected “Far” fingerprint pairings from the Miskolc dataset. This classifier performed worse than the previous one, as detailed in row 2 of Table \ref{specialized-classifiers-experiment-results}. We believe that the main cause of this performance disparity is that the data we collected from the suburban home is specially tailored to allow learning algorithms to determine the signifiers of close proximity. Because fingerprints were recorded in “bursts” of $9$ at a time, our collected data yields many more pairs of fingerprints that were recorded in the exact same location, allowing learning algorithms to more easily establish a starker contrast between “Close” and “Far” samples. The Miskolc dataset’s lack of extremely close fingerprint pairings may also be the reason why shifting the class balance slightly in favor of “Close” samples yielded better performance when using it as a training set. A secondary cause of the performance disparity may be the relative size of the training sets; the training set created from the suburban home data is more than $9$ times larger than the Miskolc dataset’s.

We also trained a second \texttt{BaggingClassifer} on the same suburban home training set, with the input feature set reduced to only the top 7 features identified by the mRMR algorithm when run on the same class-balanced data from the suburban home. Reducing the input feature set substantially reduced the time it took to train and evaluate the classifier, but reduced the classifier’s recall by $5\%$ to $6\%$, as shown in row 3 of Table \ref{specialized-classifiers-experiment-results}.

\subsection{Medium AP Density Classifier}

The median number of APs detected in a fingerprint from the TampereU and IPIN 2016 Tutorial RSSI fingerprint datasets is $38$ and $32$, respectively; thus, both datasets are of medium AP density. We created a class-balanced training set from the fingerprint pairings from the IPIN 2016 Tutorial dataset, consisting of all $30,845$ “Close” and $30,845$ randomly selected “Far” fingerprint pairings, and used it to train a \texttt{BaggingClassifier}. We evaluated the performance of that classifier on the entire set of fingerprint pairings extracted from the TampereU dataset; the results are shown in row 4 of Table \ref{specialized-classifiers-experiment-results}. We then trained an identical classifier on a training set derived from the TampereU dataset, containing all $17,193$ “Close” and $15,000$ randomly selected “Far” fingerprint pairings, and evaluated it on the full set of fingerprint pairings extracted from the IPIN 2016 Tutorial dataset. The results of this evaluation are shown in row 5 of Table \ref{specialized-classifiers-experiment-results}.

\subsection{High AP Density Classifier}
\label{high-ap-density-classifier-subsection}

The median number of APs detected in a fingerprint from the data collected from Cory Hall is between $52$ and $76$, depending on the floor, making it a high AP density environment. As detailed in Section \ref{data-sources-section}, the data from Cory Hall can be organized into a set of 9-scan “bursts,” each associated with a single physical position. A burst of fingerprints can yield data about significantly more APs than a single fingerprint; across all of the data collected from Cory Hall, the median number of APs detected in the first fingerprint of a burst is $65$, while the median number of APs detected in at least one of the first four fingerprints of a burst is $86$.

To take advantage of the additional data provided by bursts, without substantially reducing the size of the eventual training set, we divided each burst into two smaller “sub-bursts” — one comprised of the first $4$ fingerprints recorded, and the other comprised of the next $4$ fingerprints recorded. The final, ninth fingerprints were not used. For the Cory Hall datasets, the fingerprints paired up during the feature extraction process described in Section \ref{proximity-classes-section} were “pseudo-fingerprints,” one for each of the aforementioned sub-bursts. The pseudo-fingerprint for a given sub-burst contains entries for every access point detected in at least one of the sub-burst’s fingerprints. The RSSI value for a given AP in the pseudo-fingerprint is the median of all of the observed RSSI values for that AP from all of the fingerprints within the sub-burst.

We created a training set containing all $1,889$ “Close” and $300$ randomly chosen “Far” pseudo-fingerprint pairings from the first and fourth floors of Cory Hall, and used it to train a \texttt{BaggingClassifier}. We evaluated the performance of the classifier on the entire set of pseudo-fingerprint pairings from the third floor of Cory Hall. The results of this evaluation are shown in row 6 of Table \ref{specialized-classifiers-experiment-results}. The reason for choosing this particular class imbalance is that, unlike other datasets, the vast majority of the “Close” fingerprint pairings from these particular floors were recorded in the exact same place, making a classifier trained on this dataset more likely to misclassify samples recorded 1 or 2 meters apart as “Far.” As such, when a perfectly class-balanced training set is used, over $50\%$ of “Close” samples in the evaluation set are misclassified as “Far.”

For comparison, we inverted the training and evaluation process; We trained another \texttt{BaggingClassifier} on a training set derived from the pseudo-fingerprint pairings from the third floor. This training set contained all $1,549$ “Close” and $2,500$ randomly selected “Far” pseudo-fingerprint pairings. We evaluated this classifier on the full set of pseudo-fingerprint pairings from the first and fourth floors; it performed slightly better than the previous classifier, as indicated in row 7 of Table \ref{specialized-classifiers-experiment-results}. In contrast, a \texttt{BaggingClassifier} trained and evaluated on versions of the same training and evaluation sets that only used the first fingerprint of each sub-burst only achieved balanced accuracy of $65.90\%$.

\section{Conclusion and Future Work}
\label{conclusions-future-work-section}

Our experimental results show that, while a generic classifier is unable to generalize well to a wide variety of environments, an ensemble of specialized classifiers for environments with different AP densities can detect immediate physical proximity more accurately, with roughly 70\% balanced accuracy on average. Future work could explore the use of additional classifier input features, pre-processing steps, or learning algorithms to increase the classifiers’ accuracy. In addition, throughout this paper, we have manually and empirically adjusted the class balance of training sets to compensate for biases and achieve the best performance on evaluation datasets. Future work could focus on automated ways of finding the optimal class balance for training sets.

In an actual contract tracing system, participating devices would need to continuously record and save RSSI fingerprints to later periodically compare with fingerprints recorded by infected individuals. Those infected individuals’ RSSI fingerprints would need to be distributed to participating devices by centralized servers. Another practical consideration is that even if two devices are physically close to each other, their two users are not necessarily in the same room; they could be separated by a wall or door. As such, our proposed methods are merely the basic building blocks of any practical contact tracing system, and require additional enhancements before they can be deployed.

\section*{Acknowledgments}

We would like to thank Oleksii Volkovskyi and Jerome Quenum for collecting data in Cory Hall, Anderson Hansen for helping determine the design of this project, Richard Huang for implementing the RE3 algorithm from \cite{RE3-Paper}, and Willis Wang for evaluating the suitability of several RSSI fingerprint datasets for this project. Computational resources for this work, on the Microsoft Azure platform, were generously provided by Microsoft as an AI for Health grant.

\bibliographystyle{ieeetr}
\bibliography{References}

\begin{thebibliography}{10}

\bibitem{NHS-App-Impact-Paper}
C.~Wymant, L.~Ferretti, D.~Tsallis, M.~Charalambides, L.~Abeler-D\"{o}rner,
  D.~Bonsall, R.~Hinch, M.~Kendall, L.~Milsom, M.~Ayres, C.~Holmes, M.~Briers,
  and C.~Fraser, ``The epidemiological impact of the {NHS COVID-19} app,'' {\em
  Nature}, 2021.
\newblock (Pre-print from journal website).

\bibitem{TraceTogether-App-Site}
{Government of Singapore}, ``{TraceTogether},'' 2020.

\bibitem{CA-Notify-App-Site}
{California Department of Technology}, ``{CA Notify},'' 2020.

\bibitem{COVID-Watch-Arizona-App-Site}
{WeHealth}, ``{Covid Watch Arizona},'' 2020.

\bibitem{Outdoor-COVID-Transmission-Paper}
M.~Weed and A.~Foad, ``Rapid scoping review of evidence of outdoor transmission
  of {COVID-19},'' {\em medRxiv}, 2020.

\bibitem{Indoor-COVID-Transmission-Paper}
H.~Qian, T.~Miao, L.~Liu, X.~Zheng, D.~Luo, and Y.~Li, ``Indoor transmission of
  {SARS-CoV-2},'' {\em medRxiv}, 2020.

\bibitem{Indoor-Positioning-Survey-Paper}
G.~M. Mendoza-Silva, J.~Torres-Sospedra, and J.~Huerta, ``A meta-review of
  indoor positioning systems,'' {\em Sensors}, vol.~19, no.~20, 2019.

\bibitem{Inferring-WiFi-Paper}
P.~Sapiezynski, A.~Stopczynski, D.~K. Wind, J.~Leskovec, and S.~Lehmann,
  ``Inferring person-to-person proximity using wifi signals,'' {\em CoRR},
  vol.~abs/1610.04730, 2016.

\bibitem{Miskolc-Dataset-Paper}
G.~M. Mendoza-Silva, P.~Richter, J.~Torres-Sospedra, E.~S. Lohan, and
  J.~Huerta, ``Long-term wifi fingerprinting dataset for research on robust
  indoor positioning,'' {\em Data}, vol.~3, no.~1, 2018.

\bibitem{JUIndoorLoc-Dataset-Paper}
P.~Roy, C.~Chowdhury, D.~Ghosh, and S.~Bandyopadhyay, ``{JUIndoorLoc}: A
  ubiquitous framework for smartphone-based indoor localization subject to
  context and device heterogeneity,'' {\em Wirel. Pers. Commun.}, vol.~106,
  pp.~739--762, May 2019.

\bibitem{IndoorLoc-Platform-Site}
E.~Sansano, R.~Montoliu, O.~Belmonte, and J.~Torres-Sospedra, ``{UJI} indoor
  positioning and navigation repository,'' 2016.

\bibitem{Redpin-Paper}
P.~Bolliger, ``Redpin - adaptive, zero-configuration indoor localization
  through user collaboration,'' in {\em Proceedings of the First ACM
  International Workshop on Mobile Entity Localization and Tracking in GPS-Less
  Environments}, MELT '08, (San Francisco, California, USA), pp.~55--60,
  Association for Computing Machinery, 2008.

\bibitem{Redpin-Code-Repository}
P.~Brogle and P.~Bolliger, ``{Redpin} on {SourceForge}.''

\bibitem{HLF-Pair-Ratio-Paper}
M.~B. Kj\ae{}rgaard and C.~V. Munk, ``Hyperbolic location fingerprinting: A
  calibration-free solution for handling differences in signal strength
  (concise contribution),'' in {\em 2008 Sixth Annual IEEE International
  Conference on Pervasive Computing and Communications (PerCom)}, pp.~110--116,
  2008.

\bibitem{RE3-Paper}
Z.~Huang, J.~Xia, H.~Yu, Y.~Guan, and J.~Chen, ``Clustering combined indoor
  localization algorithms for crowdsourcing devices: Mining {RSSI} relative
  relationship,'' in {\em 2014 Sixth International Conference on Wireless
  Communications and Signal Processing (WCSP)}, pp.~1--6, 2014.

\bibitem{Ensemble-Survey-Paper}
O.~Sagi and L.~Rokach, ``Ensemble learning: A survey,'' {\em WIREs Data Mining
  and Knowledge Discovery}, vol.~8, no.~4, p.~e1249, 2018.

\bibitem{Attribute-Bagging-Paper}
R.~Bryll, R.~Gutierrez-Osuna, and F.~Quek, ``Attribute bagging: improving
  accuracy of classifier ensembles by using random feature subsets,'' {\em
  Pattern Recognition}, vol.~36, no.~6, pp.~1291--1302, 2003.

\bibitem{Scikit-learn-Paper}
F.~Pedregosa, G.~Varoquaux, A.~Gramfort, V.~Michel, B.~Thirion, O.~Grisel,
  M.~Blondel, P.~Prettenhofer, R.~Weiss, V.~Dubourg, J.~Vanderplas, A.~Passos,
  D.~Cournapeau, M.~Brucher, M.~Perrot, and E.~Duchesnay, ``Scikit-learn:
  Machine learning in python,'' {\em J. Mach. Learn. Res.}, vol.~12,
  pp.~2825--2830, Nov. 2011.

\bibitem{mRMR-Paper}
H.~Peng, F.~Long, and C.~Ding, ``Feature selection based on mutual information
  criteria of max-dependency, max-relevance, and min-redundancy,'' {\em IEEE
  Transactions on Pattern Analysis and Machine Intelligence}, vol.~27, no.~8,
  pp.~1226--1238, 2005.

\end{thebibliography}

\end{document}